\documentclass[runningheads]{llncs}
\usepackage[T1]{fontenc}
\usepackage{graphicx}
\usepackage{amsmath,amssymb}
\usepackage{soul}
\usepackage[ruled,vlined,linesnumbered]{algorithm2e}
\usepackage{wrapfig}
\usepackage{tikz}
\usepackage{booktabs}
\usepackage{siunitx}
\usepackage{multirow}
\usetikzlibrary{arrows.meta,positioning,calc,fit}
\DontPrintSemicolon
\SetAlgoNoEnd
\usepackage[colorlinks,
            linkcolor=blue,       
            anchorcolor=blue,
            citecolor=blue,       
            urlcolor=blue         
            ]{hyperref}

\begin{document}

\title{Efficient Solving for Dynamic Data Structure Constraint Satisfaction Problem}
\titlerunning{Efficient Solving for Dynamic Data Structure CSP}
\author{
Nanbing Li\inst{1} \and
Weijie Peng\inst{1} \and
Jin Luo\inst{1} \and
Shuai Wang\inst{2} \and
Yihui Li\inst{2} \and
Jun Fang\inst{2} \and
Yun Liang\inst{1}
}

\authorrunning{N. Li et al.}

\institute{
Peking University, Beijing, China \\
\email{nbli25@stu.pku.edu.cn, \{weijiepeng, luo-jin, ericlyun\}@pku.edu.cn} 
\and
Primarius Technologies Co., Ltd., China \\
\email{\{wangshuai, liyh, fangjun\}@primarius-tech.com}
}

\maketitle

\begin{abstract}
Functional verification plays a central role in ensuring the correctness of modern integrated circuit designs, where constrained-random verification is widely adopted to generate diverse stimuli under high-level constraints. In industrial verification environments, constraint solving increasingly involves dynamic data structures whose shape and content are determined at runtime, causing the sets of variables and constraint instances to evolve across solver invocations, which in turn leads to substantial overhead when nested and high-dimensional structures repeatedly expand across solves. We formalize this class of problems as the \emph{Dynamic Data Structure Constraint Satisfaction Problem} ($\text{D}^2\text{SCSP}$), which captures the interaction between dynamic data structure expansion and constraint evaluation. We propose a dependency-guided problem partitioning framework combined with an incremental encoding and constraint activation mechanism, enabling reuse of solver state and encodings across multiple solves. The framework is integrated into an industrial SystemVerilog verification flow and implemented in the commercial simulator VeriSim. Experimental results on industrial benchmarks demonstrate significant performance improvements, achieving an average speedup of $24.80\times$ over a baseline and $1.72\times$ over a state-of-the-art commercial simulator, highlighting the practicality of the approach for real-world verification workflows.
\end{abstract}

\section{Introduction}
Functional verification is a critical task in modern digital integrated circuit design, ensuring that a design correctly implements its specification across the intended behavior space~\cite{molina2007fv,ludden2002power4,foster2016wrg}. As design complexity increases due to microarchitectures' diversity, concurrency, and complex interaction protocols, directed testing alone becomes insufficient to achieve adequate coverage~\cite{tasiran2001coverage,jayasena2023directed}. Constrained-random verification addresses this challenge by expressing verification intent as declarative constraints over random variables, enabling systematic and scalable exploration of complex design behaviors~\cite{mehta2018,teplitsky2013coverage}. These constraints form a \emph{Constraint Satisfaction Problem} (CSP)~\cite{brailsford1999csp}, in which valid test stimuli correspond to satisfying assignments, and can be solved by built-in CSP solvers provided by SystemVerilog simulators via the \texttt{randomize()} mechanism~\cite{wu2013framework,chen2013reuse}.

In realistic verification environments, constraints and random variables can be enabled or disabled dynamically~\cite{ridgeway2012svcr}, placing randomization in the setting of \emph{Dynamic Constraint Satisfaction Problem} (DCSP). DCSP extends classical CSP to evolving constraint system sequences of related problem instances rather than a single static problem~\cite{mittal1990dcsp,wallace2009dcsp}. Prior research on DCSP has focused on reactive strategies for handling dynamic environments, including solution maintenance and reuse of reasoning or propagation to reduce redundant search~\cite{verfaillie2005,schiex1995,verfaillieschiex1994b}. Incremental arc-consistency and dynamic nogood techniques further extend static propagation and backtracking to dynamic settings by updating consistency and search state under local changes instead of recomputing from scratch~\cite{schiexverfaillie1994a,schiexverfaillie1994b}.

In more complex industrial verification scenarios, many constraint-based tasks involve dynamic data structures whose contents are determined at runtime, such as cache-coherency state tracking, transaction-level scoreboarding with dynamically sized queues, and resource allocation monitoring~\cite{cohen2023,shacham2008relaxedscoreboard,freitas2013concurrent}. These scenarios correspond to a variant of DCSP driven by data structures whose expansion is determined by assignments to specific variables, which we term the \emph{Dynamic Data Structure Constraint Satisfaction Problem} ($\text{D}^2\text{SCSP}$).

However, the challenges posed by $\text{D}^2\text{SCSP}$ are not fully addressed by existing DCSP techniques, which typically assume a fixed or gently evolving constraint scope and focus on maintaining consistency~\cite{bessiere1991arc,amilhastre2002}. In contrast, dynamic data structures induce constraint systems whose
structure and scope may expand during solving, depending on the assignments
of certain variables. This relation prevents the entire problem from being solved within a single solver invocation and instead requires decomposition into an ordered sequence of interdependent subproblems, where solutions to earlier subproblems constrain subsequent ones. In practical scenarios with nested and high-dimensional data structures, such expansion quickly introduces complex dependencies that undermine locality assumptions and lead to rapid growth in both problem size and solving time.

To address these challenges, we introduce a modeling framework based on a refined variable representation that captures dependency relations and expansion semantics induced by dynamic data structures. Guided by these dependencies, the problem is decomposed into a sequence of subproblems, where incremental solving is applied at the granularity of individual subproblems. In industrial practice, CSPs are commonly encoded into Satisfiability formulations~\cite{davidson2020cp2smt,bofill2020rcpsp} and solved using assumption-based incremental solving in modern solvers~\cite{improvingglucose2016,cimatti2012}, allowing both encoding and solver state to be reused across successive solves.


This paper introduces the framework for efficient dynamic data structure constraint solving through the following key contributions:
\begin{enumerate}
  \item \textbf{A modeling framework for Dynamic Data Structure Constraint Satisfaction Problems ($\text{D}^2\text{SCSP}$)} that abstracts dependency relations in dynamic data structures and clarifies semantics to support efficient solving.
  
  \item \textbf{A dependency-guided problem partitioning approach for $\text{D}^2\text{SCSP}$} that derives a solving order leveraging constraint locality and maximizing joint solving scope to reduce solving overhead.

  \item \textbf{An incremental encoding and solving mechanism} that reduces redundant work across iterations by selectively activating constraint fragments affected by runtime changes and reusing solver knowledge, such as learned clauses across repeated solves.
  
\end{enumerate}

We integrate this architecture into an industrial verification flow, realizing a practical tool that supports dynamic constraint scenarios without requiring modifications to existing specifications. Implemented within the commercial simulator, VeriSim~\cite{primariusverisim_web}, for SystemVerilog-based constrained-random verification, our approach delivers substantial performance gains, achieving up to $24.80\times$ speedup over the original VeriSim implementation and $1.72\times$ speedup compared to the state-of-the-art commercial tool, Synopsys VCS~\cite{synopsysvcsT2022ucli}.
\section{Preliminaries}

\subsection{Variables, Assignments, and Static Constraints}

Let $V=\{v_1,\dots,v_n\}$ be a finite set of variables, where each
$v\in V$ ranges over a finite domain $Dom(v)$. An \emph{assignment} is an
element $\sigma\in\prod_{v\in V} Dom(v)$. For any $V'\subseteq V$, the
projection of $\sigma$ onto $V'$, denoted by $\pi(\sigma,V')$, is given by
$(\sigma(v))_{v\in V'}$. A \emph{static constraint} $c$ over a variable
set $V_c\subseteq V$ is a relation
$c\subseteq\prod_{v\in V_c} Dom(v)$, where $V_c$ is referred to as the
\emph{scope} of $c$. An assignment $\sigma$ satisfies $c$, written
$\sigma\models c$, if $\pi(\sigma,V_c)\in c$. Given a finite set of
constraints $C$, an assignment $\sigma$ satisfies $C$, written
$\sigma\models C$, if $\sigma\models c$ for all $c\in C$.

\subsection{Constraint Satisfaction Problem}

A \emph{Constraint Satisfaction Problem} (CSP) is defined as a triple
$\langle V, Dom, C \rangle$, where $V$ is a finite set of variables, $Dom$
assigns a finite domain $Dom(v)$ to each $v \in V$, and $C$ is a finite set
of static constraints over $V$. A solution to a CSP is an assignment
$\sigma \in \prod_{v\in V} Dom(v)$ such that $\sigma \models C$.

A \emph{Dynamic Constraint Satisfaction Problem} (DCSP) is defined as a
sequence of related constraint satisfaction problems
$\langle (V_0,Dom_0,C_0),\dots,(V_T,Dom_T,C_T)\rangle$, where each
$(V_i,Dom_i,C_i)$ is a static CSP and the variable, domain, and constraint sets may change between successive instances. A solution to a
DCSP is a sequence of assignments
$\langle \sigma_0,\dots,\sigma_T\rangle$ such that for each $i$,
$\sigma_i \in \prod_{v\in V_i} Dom_i(v)$ and $\sigma_i \models C_i$.

\section{Modeling}

\subsection{Dynamic Data Structures and Dummy Variables}

A \emph{dynamic data structure} is a mapping type
$\mathcal{D} : \mathcal{K} \to \mathcal{V}$, where $\mathcal{K}$ is a
finite key domain and $\mathcal{V}$ is a finite value domain. Each
dynamic data structure type $\mathcal{D} : \mathcal{K} \to \mathcal{V}$
is associated with a corresponding \emph{dummy variable}
$\mathit{dv}\langle \mathcal{K}, \mathcal{V} \rangle$, whose domain $Dom(\mathit{dv}\langle \mathcal{K}, \mathcal{V} \rangle)
\triangleq \mathcal{K} \to \mathcal{V}$. Once a finite subset $\mathcal{K}^\ast \subseteq \mathcal{K}$ is
determined, the \emph{expansion} operator
$\mathsf{expand}(\mathit{dv}, \mathcal{K}^\ast)$ maps
$\mathit{dv}\langle \mathcal{K}, \mathcal{V} \rangle$ to a finite set of
indexed variables $\{\mathit{dv}[k] \mid k \in \mathcal{K}^\ast\}$, where
each $\mathit{dv}[k]$ denotes the variable corresponding to key $k$ with
value domain $\mathcal{V}$.

The dynamic nature of a data structure is captured by the fact that the
effective key set used for expansion is not fixed, but is determined by
the values of a finite set of variables. Formally, for each dynamic data
structure type $\mathcal{D} : \mathcal{K} \to \mathcal{V}$, there exists
a function $f_{\mathcal{D}} : \prod_{i=1}^{n} Dom(x_i) \to 2^{\mathcal{K}}$,
where $(x_1,\dots,x_n) \subseteq V$. Given an assignment $\sigma$, the key
set used for expansion is
$\mathcal{K}^\ast \triangleq
f_{\mathcal{D}}(\sigma(x_1),\dots,\sigma(x_n))$. Accordingly, the
expansion of the dummy variable
$\mathit{dv}\langle \mathcal{K}, \mathcal{V} \rangle$ is enabled only
after the values of the variables $x_1,\dots,x_n$ are determined, and
different assignments may induce different expansion scopes.

\begin{wrapfigure}[11]{r}{0.73\textwidth} 
\centering
  \vspace{-15pt}                
  \begin{tikzpicture}[
  font=\footnotesize,
  >=Latex,
  expand_solid/.style={->, line width=0.5pt, solid}, 
  path_dashed/.style={->, dashed, line width=0.5pt}, 
  dep/.style={->, dashed, line width=0.7pt},
  dv/.style ={draw, rounded corners=2pt, inner sep=2pt, minimum height=6mm, fill=white},
  var/.style={draw, circle, inner sep=1pt, minimum size=10mm, fill=white},
  box/.style={draw, rounded corners=2pt, inner sep=2pt, minimum height=6mm, fill=white},
  hint/.style={font=\tiny}
]

\def\colOne{-5.5}     
\def\colOneHalf{-3.5} 
\def\colTwo{-1.2}     
\def\colThree{1.8}    

\node[var] (root) at (\colOne, 0) {$\mathit{arr.size}$};
\node[hint, below=1pt of root] {$\sigma(\mathit{x})=2$};

\node[dv] (dv_arr) at (\colOneHalf, 1.2) {$\mathit{arr[][]}$};
\node[dv] (dv_size) at (\colOneHalf, -1.2) {$\mathit{arr[].size}$};

\draw[path_dashed] (root.north east) -- (dv_arr.west);
\draw[path_dashed] (root.south east) -- (dv_size.west);

\node[dv] (a0) at (\colTwo, 1.55) {$\mathit{arr}[0][]$};
\node[box] (s0) at (\colTwo, 0.5) {$\mathit{arr}[0].size$};
\node[hint, below=0.3pt of s0] {$\sigma(\mathit{y_{0}})=2$};

\draw[expand_solid] (dv_arr.east) -- (a0.west);
\draw[expand_solid] (dv_size.east) -- (s0.west); 
\draw[dep] (s0) -- (a0); 

\node[hint, anchor=south, inner sep=1pt] at ($(dv_arr.east)!0.5!(a0.west)$) {0};
\node[hint, anchor=north, inner sep=1pt] at ($(dv_size.east)!0.5!(s0.west)$) {0};

\node[box] (e00) at (\colThree, 1.55) {$\mathit{arr}[0][0]$};
\node[box] (e01) at (\colThree, 0.5) {$\mathit{arr}[0][1]$};
\draw[expand_solid] (a0.east) -- node[hint, above, pos=0.5] {0} (e00.west);
\draw[expand_solid] (a0.east) -- node[hint, below, pos=0.5] {1} (e01.west);

\node[dv] (a1) at (\colTwo, -0.5) {$\mathit{arr}[1][]$};
\node[box] (s1) at (\colTwo, -1.55) {$\mathit{arr}[1].size$};
\node[hint, below=0.5pt of s1] {$\sigma(\mathit{y_{1}})=2$};

\draw[expand_solid] (dv_arr.east) -- (a1.west);
\draw[expand_solid] (dv_size.east) -- (s1.west); 
\draw[dep] (s1) -- (a1);

\node[hint, anchor=south, inner sep=1pt] at ($(dv_arr.east)!0.5!(a1.west)$) {1};
\node[hint, anchor=north, inner sep=1pt] at ($(dv_size.east)!0.5!(s1.west)$) {1};

\node[box] (e10) at (\colThree, -0.5) {$\mathit{arr}[1][0]$};
\node[box] (e11) at (\colThree, -1.55) {$\mathit{arr}[1][1]$};
\draw[expand_solid] (a1.east) -- node[hint, above, pos=0.5] {0} (e10.west);
\draw[expand_solid] (a1.east) -- node[hint, below, pos=0.5] {1} (e11.west);

\end{tikzpicture}
  \vspace{-15pt}                
  \caption{Illustrative expansion for a 2D dynamic array}
  \vspace{-25pt}                
  \label{fig:arr-expansion-bottom}
\end{wrapfigure}
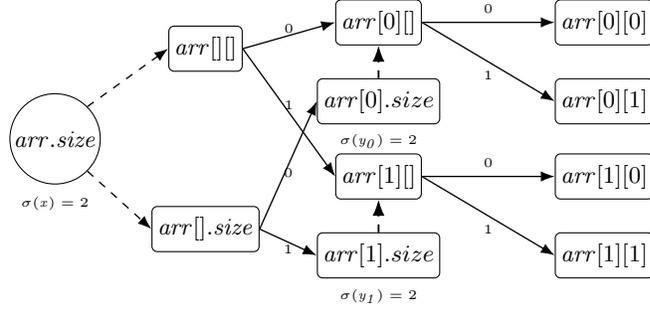

As illustrated in Fig.~\ref{fig:arr-expansion-bottom}, consider a
two-dimensional dynamic array $\mathit{arr}[][]$, modeled as a dynamic
data structure of type $\mathcal{D} : \mathcal{K} \to (\mathcal{K} \to
\mathcal{V})$ with dummy variable
$\mathit{dv}_{\mathit{arr}}\langle \mathcal{K}, \mathcal{K}\!\to\!\mathcal{V} \rangle$.
Let $x=\mathit{arr.size}$ be the variable determining the first-dimension
index set, and let $f_{\mathcal{D}}$ be the corresponding key-generation
function. Given an assignment $\sigma$ with $\sigma(x)=2$, we obtain
$\mathcal{K}^\ast = f_{\mathcal{D}}(\sigma(x)) = \{0,1\}$, and
$\mathsf{expand}(\mathit{dv}_{\mathit{arr}}, \{0,1\})$ yields the
dummy variables $\mathit{arr}[0][]$ and $\mathit{arr}[1][]$. Concurrently,
$\mathsf{expand}(\mathit{dv}_{\mathit{arr[].size}}, \{0,1\})$ produces the
variables $\mathit{arr}[0].size$ and $\mathit{arr}[1].size$. For each
$i\in\{0,1\}$, let $y_i=\mathit{arr}[i].size$ determine the
second-dimension index set via the corresponding function
$f_{\mathcal{D}}$. If $\sigma(y_i)=2$, then
$f_{\mathcal{D}}(\sigma(y_i))=\{0,1\}$, and
$\mathsf{expand}(\mathit{dv}_{\mathit{arr}[i][]}, \{0,1\})$ yields the
variables $\mathit{arr}[i][0]$ and $\mathit{arr}[i][1]$.

\subsection{Dynamic Constraints}

A \emph{dynamic constraint} $\mathrm{DC}(V_d,D_d)$ over a variable set
$V_d \cup D_d$, where $V_d$ is a finite set of variables and
$D_d$ is a finite set of dummy variables, is a relation
$\mathrm{DC}(V_d,D_d) \subseteq
\prod_{x \in V_d \cup D_d} Dom(x)$, and $V_d \cup D_d$ is
referred to as the \emph{scope} of the dynamic constraint.

The \emph{expansion} of a dynamic constraint is defined with respect to a
chosen dummy variable $\mathit{dv}\in D_d$ and a finite key set
$\mathcal{K}^\ast \subseteq \mathcal{K}_{\mathit{dv}}$. The expansion of
$\mathrm{DC}(V_d,D_d)$ with respect to
$(\mathit{dv},\mathcal{K}^\ast)$ is a finite set of dynamic constraints
$\mathsf{expand}\allowbreak
(\mathrm{DC}(V_d,D_d),\mathit{dv},\mathcal{K}^\ast)
\mathrel{\triangleq}
\{\mathrm{DC}(V_d^{k},D_d^{k}) \mid k\in\mathcal{K}^\ast\}$,
where for each $k\in\mathcal{K}^\ast$,
$V_d^{k} \triangleq V_d \cup \{\mathit{dv}[k]\}$
and
$D_d^{k} \triangleq (D_d \setminus \{\mathit{dv}\})
\cup \{\mathit{dv}[k] \mid Dom(\mathit{dv}[k]) = \mathcal{K}' \to \mathcal{V}'\}$.
When $D_d^{k}=\emptyset$, the dynamic constraint
$\mathrm{DC}(V_d^{k},D_d^{k})$ degenerates into a static constraint,
i.e., a relation $c^{k} \subseteq \prod_{v\in V_d^{k}} Dom(v)$.

\subsection{Dynamic Data Structure Constraint Satisfaction Problem}

Given a finite set of variables $V$ and a finite set of dummy variables
$D$, a \emph{Dynamic
Data Structure Constraint Satisfaction Problem} ($\text{D}^2\text{SCSP}$) is defined by a tuple
$\langle V,D,C_s,C_d\rangle$, where $C_s$ is a finite set of
static constraints over $V$, and $C_d$ is a finite set of dynamic
constraints of the form $\mathrm{DC}(V_d,D_d)$ with
$V_d \subseteq V$ and $D_d \subseteq D$, without explicitly modeling domain evolution.

A \emph{solution} to a $\text{D}^2\text{SCSP}$ instance $\langle V,D,C_s,C_d\rangle$ is a total assignment $\sigma_T \in \prod_{v\in V_T} Dom(v)$ obtained as the final state of a finite sequence
 $\langle
(V_0,\allowbreak D_0,\allowbreak C_s^0,\allowbreak C_d^0,\allowbreak \sigma_0),
\allowbreak \dots,\allowbreak
(V_T,\allowbreak D_T,\allowbreak C_s^T,\allowbreak C_d^T,\allowbreak \sigma_T)
\rangle$,
where $V_0=V$, $D_0=D$, $C_s^0=C_s$, $C_d^0=C_d$, each $\sigma_i \in \prod_{v\in V_i} Dom(v)$, and $\pi(\sigma_{i+1},V_i)=\sigma_i$ for all $i<T$.
For each state $i<T$, there exist $\mathrm{DC}(V_d,D_d)\in C_d^i$, $\mathit{dv}\in D_d$, and a finite key set $\mathcal{K}^\ast \triangleq f_{D}(\sigma_i(x_1),\dots,\sigma_i(x_n))$ such that
$\mathsf{expand}(\mathrm{DC}(V_d,D_d),\mathit{dv},\mathcal{K}^\ast)=\{\mathrm{DC}(V_d^{k},D_d^{k}) \mid k\in\mathcal{K}^\ast\}$.
The system state evolves by materializing all expansions, yielding
$V_{i+1}=V_i \cup \bigcup_{k\in\mathcal{K}^\ast}(V_d^{k}\setminus V_d)$,
$D_{i+1}=(D_i\setminus\{\mathit{dv}\})\cup\bigcup_{k\in\mathcal{K}^\ast}D_d^{k}$,
$C_d^{i+1}=(C_d^i\setminus\{\mathrm{DC}(V_d,D_d)\})\cup\{\mathrm{DC}(V_d^{k},D_d^{k}) \mid D_d^{k}\neq\emptyset\}$,
and $C_s^{i+1}=C_s^i\cup\{c^{k}\mid D_d^{k}=\emptyset\}$.
The sequence terminates when $D_T=\emptyset$ and $C_d^T=\emptyset$, and the final assignment satisfies $\sigma_T\models C_s^T$.
The objective is to generate a set $\{\sigma_T^{(1)},\dots,\sigma_T^{(N)}\}$ of such final assignments together with their corresponding generating sequences.

\section{Solving Strategy}
\subsection{Problem Partitioning}

In classical CSP, all variables and constraints can be submitted to the solver at once, and the order in which variables are assigned is a heuristic that affects solving efficiency~\cite{song2021learning}.
This assumption no longer holds in $\text{D}^2\text{SCSP}$, where assignments to certain variables induce dynamic expansion of variables and constraints during solving.
Consequently, $\text{D}^2\text{SCSP}$ requires a dependency-guided problem partitioning, in which the constraint system is decomposed into an ordered sequence of subproblems.

This partitioning arises from the following two relations.

\textit{Dependency.}
Dynamic expansion semantics induce a dependency relation over variables and dummy variables.
Let $\mathit{dv}\langle \mathcal{K}, \mathcal{V} \rangle \in D$ be a dummy variable whose expansion key set is generated by a function
$f_{D} : \prod_{i=1}^{n} Dom(x_i) \to 2^{\mathcal K}$ for a finite set of variables
$(x_1,\dots,x_n)\subseteq V$.
For each $i\in\{1,\dots,n\}$, we define a dependency relation
$x_i \prec \mathit{dv}$,
indicating that the value of $x_i$ must be fixed before the expansion of $\mathit{dv}$ is well-defined.

\textit{Co-occurrence.}
The co-occurrence relation captures variables that appear within the same constraint scope.
For a static constraint $c$ over $V_c \subseteq V$, we define $u \sim v$ for all $u,v \in V_c$.
For a dynamic constraint $\mathrm{DC}(V_d,D_d)$, we define $u \sim v$ for all
$u,v \in V_d \cup D_d$.
Variables related by $\sim$ are preferably solved in the same stage to preserve constraint locality.

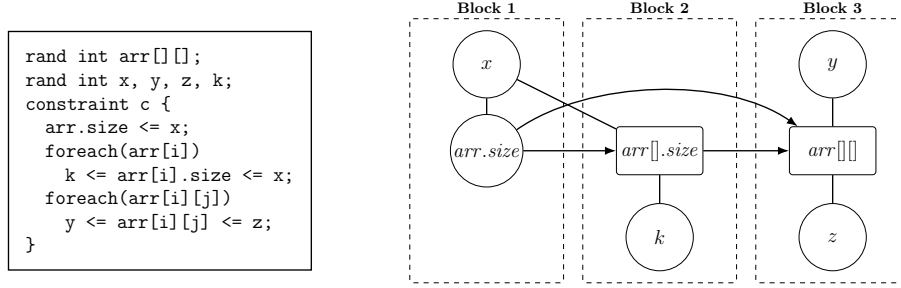
\begin{figure}[t]
\centering
\resizebox{\textwidth}{!}{
\begin{tikzpicture}[
    font=\small,
    >=Latex,
    block/.style={draw, dashed, black, rounded corners=2pt, 
                  minimum width=3.2cm, minimum height=5.5cm},
    node_var/.style={draw, circle, minimum size=14mm, fill=white, inner sep=1pt, font=\large},
    node_dv/.style={draw, rectangle, rounded corners=2pt, minimum width=18mm, 
                   minimum height=10mm, fill=white, inner sep=2pt, font=\large},
    edge_dep/.style={->, thick}, 
    edge_co/.style={-, thick},   
    label_blk/.style={font=\bfseries, black}
]

\node[anchor=center, align=left, draw, thick, fill=white, inner sep=10pt, font=\large] (code) at (-6.8, 0) {
    \texttt{rand int arr[][];} \\
    \texttt{rand int x, y, z, k;} \\
    \texttt{constraint c \{} \\
    \quad \texttt{arr.size <= x;} \\
    \quad \texttt{foreach(arr[i])} \\
    \quad \quad \texttt{k <= arr[i].size <= x;} \\
    \quad \texttt{foreach(arr[i][j])} \\
    \quad \quad \texttt{y <= arr[i][j] <= z;} \\
    \texttt{\}}
};

\def\baseY{0}
\def\colGap{3.6} 

\node[block] (B1) at (0, \baseY) {};
\node[label_blk, anchor=south] at (B1.north) {Block 1};
\node[node_var] (asize) at (0, 0) {$\mathit{arr.size}$};
\node[node_var] (x) at (0, 1.8) {$x$}; 

\node[block] (B2) at (\colGap, \baseY) {};
\node[label_blk, anchor=south] at (B2.north) {Block 2};
\node[node_dv] (isize) at (\colGap, 0) {$\mathit{arr[ ].size}$};
\node[node_var] (k) at (\colGap, -1.8) {$k$}; 

\node[block] (B3) at (2*\colGap, \baseY) {};
\node[label_blk, anchor=south] at (B3.north) {Block 3};
\node[node_dv] (arr_dv) at (2*\colGap, 0) {$\mathit{arr[ ][ ]}$};
\node[node_var] (y) at (2*\colGap, 1.8) {$y$};  
\node[node_var] (z) at (2*\colGap, -1.8) {$z$}; 


\draw[edge_dep] (asize) -- (isize);
\draw[edge_dep] (isize) -- (arr_dv);

\draw[edge_co] (asize) -- (x);

\draw[edge_co] (isize) -- (x); 
\draw[edge_co] (isize) -- (k);

\draw[edge_co] (arr_dv) -- (y);
\draw[edge_co] (arr_dv) -- (z);

\draw[edge_dep] (asize) to [out=35, in=145, looseness=0.8] (arr_dv);

\end{tikzpicture}
}
\caption{Dynamic constraint graph partitioned into ordered solving blocks.}
\label{fig:DSCSP-graph-blocks-k}
\end{figure}

As in CSP, where constraint structures are commonly represented as hypergraphs with each constraint forming a hyperedge~\cite{marx2009hypergraph},
we derive a dynamic constraint graph through a primal mixed graph construction over variables and dummy variables.
 In the dynamic constraint graph, the co-occurrence relation $u \sim v$ is represented as an undirected edge,
  and the dependency relation $u \prec v$ is represented as a directed edge. Fig.~\ref{fig:DSCSP-graph-blocks-k} shows the dynamic constraint graph over variables and
dummy variables.
Dummy-variable nodes represent families of variables whose concrete instances are
created only after runtime expansion; the node $\mathit{arr[][]}$ abstracts both the
intermediate dummy-variable set $\{\mathit{arr}[i][]\}$ and concrete variables
$\{\mathit{arr}[i][j]\}$ generated by subsequent expansions.

\begin{algorithm}[t]
\caption{\textsc{PartitionGraph}$(V,D,\prec,\sim)$}
\label{alg:partition-graph}
$U \gets V \cup D$\;
$\mathcal X \gets$ the set of connected components of $(U,\sim)$\;
\ForEach{$\Xi\in\mathcal X$}{
  initialize $\mathsf{level}(x)\gets false$ for all $x \in \Xi$\;
  assign $\mathsf{level}(\cdot)$ by a BFS over $\prec$ restricted to $\Xi$\;
  \ForEach{$x\in \Xi$}{
    $R(x)\gets \{y\in \Xi \mid y \text{ is reachable from } x \text{ via } \sim\}$\;
    $\ell \gets \min\{\mathsf{level}(d)\mid d\in R(x)\cap  D \land \mathsf{level}(d)\neq false\}$\;
    $\mathsf{level}(x)\gets (\ell \text{ defined})\ ?\ \ell\ :\ 0$\;
  }
}
$K\gets \max\{\mathsf{level}(x)\mid x \in U\}$\;
\For{$i\gets 0$ \textbf{to} $K$}{
  $B_i\gets\{x\in U \mid \mathsf{level}(x)=i\}$\;
}
\KwRet $\langle B_0,B_1,\dots,B_K\rangle$\;
\end{algorithm}

Since the dynamic constraint graph may consist of multiple disconnected components, each component can be handled independently. Within each component, as illustrated in Fig.~\ref{fig:DSCSP-graph-blocks-k}, the dynamic constraint graph is partitioned into an ordered sequence of solving blocks, corresponding to a decomposition of the original problem into subproblems solved sequentially, where all directed edges go from earlier to later blocks and dependency relations are respected. After variables in a block are solved, their assignments are fixed and propagated as constants to subsequent blocks, reducing their domains to singletons, while variables in earlier blocks are solved without considering constraints from later blocks, potentially restricting feasible extensions due to premature fixing. To reduce the overhead caused by solving failures and subsequent re-solving, we aim to place variables connected by undirected edges into the same block whenever possible, allowing them to be solved jointly with larger effective domains.

Algorithm~\ref{alg:partition-graph} partitions each connected component $\Xi$ independently and then merges the resulting levels into a global ordered partition. Within a component, it first assigns levels by a BFS traversal over the dependency relation $\prec$ (lines~5), and then places the remaining elements using the co-occurrence relation $\sim$ by assigning them to the minimum feasible level (lines~6-9). This encourages constraints involving these elements to be enforced earlier, when more related variables remain unfixed, avoiding overly restricted domains caused by premature fixing across blocks. The algorithm yields a sequence of solving blocks $\langle B_0,\dots,B_K\rangle$, where $B_i=\{x\in V\cup D \mid \mathsf{level}(x)=i\}$. Each constraint instance $c$ is associated with block $B_{\max\{\,\mathsf{level}(x)\mid x \text{ occurs in } c\,\}}$.

\subsection{Incremental Encoding and Solving}

Constraint satisfaction problems can be effectively solved through encoding and solving in \emph{Satisfiability} (SAT)~\cite{petke2015}.
However, in $\text{D}^2\text{SCSP}$, runtime-determined expansions can blow up the number of instantiated variables and constraints, making SAT encoding and solving costly.
At the same time, successive expansions exhibit substantial structural reuse: previously materialized instances are retained, and only keys in $\Delta\mathcal K^\ast$ introduce new variables and constraints, leaving room for incremental encoding that focuses on $\mathsf{expand}(\mathit{dv},\Delta\mathcal K^\ast)$.

Given a $\text{D}^2\text{SCSP}$ instance $\langle V,D,C_s,C_d\rangle$,
we associate each instantiated constraint with a Boolean guard literal.
For any dynamic constraint $\mathrm{DC}(V_d,D_d)\in C_d$, we define its
guarded form as $(\mathrm{DC}(V_d,D_d), g_0)$ with $g_0 \triangleq true$.
Let $\{h_k \mid k\in\mathcal K\}$ be a set of fresh Boolean guard literals.
For a guarded dynamic constraint $(\mathrm{DC}(V_d,D_d), g)$ and a chosen
dummy variable $\mathit{dv}\in D_d$ with effective key set
$\mathcal K^\ast \subseteq \mathcal K_{\mathit{dv}}$, we define the guarded expansion as
\begin{equation*}
\label{eq:guarded-expand}
\mathsf{expand}_g\!\left(\mathrm{DC}(V_d,D_d), g, \mathit{dv}, \mathcal K^\ast\right)
\triangleq
\left\{
\left(\mathrm{DC}(V_d^{k},D_d^{k}),\; g\land h_k\right)
\ \middle|\ k\in\mathcal K^\ast
\right\}.
\end{equation*}
Along an expansion path selecting keys $(k_1,k_2,\dots)$, the corresponding guards satisfy
$g_{k_0}\triangleq g_0$ and
$g_{k_i}\triangleq g_{k_{i-1}}\land h_{k_i}$ for all $i\ge 1$.
If $D_d^{k_i}=\emptyset$, the resulting instance degenerates to a static constraint
$c^{k_i}$ and is encoded as
$g_{k_i}\Rightarrow \mathsf{Enc}(c^{k_i})$.

We consider repeated solving of a $\text{D}^2\text{SCSP}$ instance.
Let $\mathcal K_{t-1}^\ast$ and $\mathcal K_t^\ast$ denote the effective key sets
at solves $t{-}1$ and $t$, respectively.
A guarded constraint instance with guard $g_{k_i}=false$ is disabled and does not
participate in further expansion, while only instances with $g_{k_i}=true$
remain active.
Keys in $\mathcal K_t^\ast \cap \mathcal K_{t-1}^\ast$ reuse previously
materialized expansion results and do not trigger additional instantiation.
For a dynamic constraint $\mathrm{DC}(V_d,D_d)$, the temporal expansion
under an active guard is defined as
{\small
\begin{equation*}
\label{eq:expand-temporal}
\mathsf{expand}_g\!\left(
\mathrm{DC}(V_d,D_d), true,\mathit{dv},
\mathcal K_{t-1}^\ast,\,\mathcal K_t^\ast
\right)
\triangleq
\begin{cases}
\begin{aligned}
\bigl(\mathrm{DC}(V_d^{k_i},D_d^{k_i}),\, true\bigr),\\
\quad k_i \in \mathcal K_t^\ast \setminus \mathcal K_{t-1}^\ast,
\end{aligned}
\\[6pt]
\begin{aligned}
\bigl(\mathrm{DC}(V_d^{k_i},D_d^{k_i}),\, false\bigr),\\
\quad k_i \in \mathcal K_{t-1}^\ast \setminus \mathcal K_t^\ast.
\end{aligned}
\end{cases}
\end{equation*}
}
For each $k_i$ such that $D_d^{k_i}=\emptyset$, the static constraint
$c^{k_i}$ is encoded in guarded form as
$g_{k_i} \Rightarrow \mathsf{Enc}(c^{k_i})$.
Let $\Gamma^{(t)}$ denote the encoding set after the $t$-th solve; it grows as
$\Gamma^{(t)}
=
\Gamma^{(t-1)} \cup
\{\, g_{k_i} \Rightarrow \mathsf{Enc}(c^{k_i})
\mid k_i \in \mathcal K_t^\ast \land D_d^{k_i}=\emptyset \,\}$.

Modern CDCL-based SAT solvers support incremental solving under assumptions, enabling repeated solver invocations while reusing learned clauses, heuristics, and other internal state~\cite{cimatti2012}. In our setting, each materialized constraint instance $c^{k_i}$ is encoded into SAT in guarded form as
$g_{k_i} \Rightarrow \mathsf{Enc}(c^{k_i})$, which is equivalent to the clause $(\lnot g_{k_i} \lor \mathsf{Enc}(c^{k_i}))$, where $g_{k_i}$ is a fresh Boolean guard variable. Guard variables are handled via the SAT solver's assumption mechanism: each guarded clause is added once to the solver, and successive solves differ only in the assumptions, where $g_{k_i}=\mathsf{true}$ enforces $\mathsf{Enc}(c^{k_i})$ and $g_{k_i}=\mathsf{false}$ makes the clause trivially satisfied. In addition, assignments produced by earlier solving blocks are passed as assumptions in subsequent solves, thereby fixing their values. Across repeated solves of the same $\text{D}^2\text{SCSP}$ instance, the underlying SAT formula—consisting of all guarded clauses—remains fixed, and only the assumption set over guard variables changes, enabling incremental SAT solving with full reuse of learned clauses and solver state.

Beyond SAT, CSP can also be solved by encoding them into \emph{Satisfiability Modulo Theories} (SMT)~\cite{davidson2020cp2smt,bofill2009}.
Incremental SMT solvers use a stack-based \texttt{push()}/\texttt{pop()} mechanism to scope assertions and retract them as needed~\cite{bembenek2020datalog}.

We maintain a cached set of assumption encodings for each solving block $B$.
For the $t$-th solve under $B$, let $\Lambda_B^{(t)}$ denote the set of encodings corresponding to all assumptions active in $B$ at this solve.
For block $B$, a local solving context is created by a single \texttt{push} upon its first solve, on top of the solver's native assumption-based incremental interface.
If $\Lambda_B^{(t)}=\Lambda_B^{(t-1)}$, the solver is invoked directly under the same pushed context.
Otherwise, the cached context is discarded by a corresponding \texttt{pop}, and a fresh \texttt{push} is issued to assert the updated set $\Lambda_B^{(t)}$.
Permanent assertions invalidate the cache and are added outside the pushed frame, ensuring that only per-solve assumption encodings are scoped by \texttt{push}/\texttt{pop} while all other solver state is preserved.

\section{Implementation}

The proposed $\text{D}^2\text{SCSP}$ mechanism is implemented by extending the standard \texttt{randomize()} flow of the VeriSim SystemVerilog simulator~\cite{primariusverisim_web}.

\textit{Preparing.}
At the beginning of \texttt{randomize()}, the solver collects the object’s random variables and constraints and constructs an ordered sequence of solving blocks using dependency-guided partitioning; each block is bound to a persistent SMT solver instance to enable reuse of solver state and cached encodings across repeated invocations on the same object. Dynamic data structures are represented via key-dependent dummy variables that collect runtime-determined keys, together with constraint templates defining their structure. The implementation targets dynamic data structures commonly used in practice, including unpacked arrays, associative arrays, queues, and class-based objects with nested fields.

\textit{Solving.}
Solving proceeds block by block using an incremental encoding and constraint activation mechanism. Constraint expressions are encoded once as SMT bit-vector terms, assignments from earlier blocks are fixed and propagated, and only constraints whose guards become determined are activated. When solving determines new sizes or keys of dynamic data structures, the corresponding variables and constraints are materialized and scheduled into later blocks.

\textit{Commit/Rollback.} If all blocks are solved, assignments are committed; otherwise, solver rolls back and retries, reporting failure only if no solution is found.

\section{Evaluation}

All experiments were conducted on a server with dual Intel Xeon Gold 6348 CPUs (2.60 GHz, 56 cores total) and 2 TB of RAM.
We evaluate the proposed approach on real-world constrained-random verification benchmarks derived from proprietary industrial designs. The accumulated CNF encodings involved in these benchmarks average 30,705 variables and 51,143 clauses, with peaks of 95,000 variables and 157,693 clauses.
The approach is implemented in VeriSim~1.24.2~\cite{primariusverisim_web} and compared against the original VeriSim~1.24.2 and Synopsys VCS~2022.06~\cite{synopsysvcsT2022ucli}.
All experiments are executed in single-threaded mode, and both VeriSim configurations use the same SMT solver, Bitwuzla~\cite{bitwuzla2023}.
For each benchmark, 1000 invocations of the \texttt{randomize()} method are applied to the same SystemVerilog object instance and averaged over 10 runs, with reported runtimes measuring the total end-to-end compilation and simulation time.
\begin{table}[htbp]
\centering
\caption{Performance comparison of our approach (Ours), Synopsys VCS, and original VeriSim. The best runtimes are highlighted in \textbf{bold}. ``T/O'' indicates a timeout.}
\label{tab:performance_comparison}
\begin{tabular}{l|ccc|cc}
\hline
\textbf{Benchmark} & Ours (s) & VCS (s) & VeriSim (s) & over VCS & over VeriSim \\ \hline
multidim\_32b       & \textbf{8.80}  & 9.30  & 503.45  & 1.06$\times$ & 57.21$\times$ \\
multidim\_nested    & \textbf{8.98}  & 9.28  & 528.81  & 1.03$\times$ & 58.90$\times$ \\
multidim\_47b       & 10.15          & \textbf{9.92}  & 558.13  & 0.98$\times$ & 54.97$\times$ \\
unique\_simple      & \textbf{11.55} & 12.26 & 565.75  & 1.06$\times$ & 48.97$\times$ \\
unique\_nested      & 12.07          & \textbf{12.01} & 549.67  & 0.99$\times$ & 45.53$\times$ \\
bit\_count          & \textbf{1.41}  & 2.48  & 26.08   & 1.76$\times$ & 18.54$\times$ \\
mem\_alloc          & \textbf{0.64}  & 1.43  & 4.70    & 2.22$\times$ & 7.29$\times$ \\
clog                & \textbf{0.67}  & 1.18  & 5.40    & 1.78$\times$ & 8.10$\times$ \\
hash\_inverse       & \textbf{0.46}  & 1.20  & 1.34    & 2.60$\times$ & 2.91$\times$ \\

float\_compare      & \textbf{0.58}  & 1.18  & 3.78    & 2.04$\times$ & 6.53$\times$ \\
conditional\_a      & \textbf{0.55}  & 1.14  & 1.29    & 2.09$\times$ & 2.36$\times$ \\
conditional\_b      & \textbf{0.67}  & 1.41  & 2.66    & 2.12$\times$ & 4.00$\times$ \\
ite                 & \textbf{5.41}  & 15.03 & 68.69   & 2.78$\times$ & 12.71$\times$ \\
subset\_sum         & \textbf{0.84}  & 1.32  & 16.08   & 1.57$\times$ & 19.14$\times$ \\
solve\_priority     & \textbf{79.54} & T/O   & 2536.68 & --           & 31.89$\times$ \\ \hline
\textbf{Average}   & \textbf{4.48}  & 5.65  & 202.56  & \textbf{1.72$\times$} & \textbf{24.80$\times$} \\ \hline
\end{tabular}
\end{table}

Table~\ref{tab:performance_comparison} shows that the proposed approach consistently outperforms the baseline VeriSim and achieves competitive or better performance than VCS on industrial constrained-random verification benchmarks. The largest gains over VeriSim arise on benchmarks with deeply nested dynamic data structures due to incremental solving, while for constraint types less amenable to SMT encoding, VCS exhibits smaller performance gaps owing to specialized solving mechanisms.
\begin{table}[htbp]

\centering
\caption{Detailed comparison of Preprocess and Solve Time across configurations}
\label{tab:detailed_performance_reordered}
\setlength{\tabcolsep}{4pt}
\begin{tabular}{l|ccc|ccc}
\hline
\multirow{2}{*}{\textbf{Benchmark}} & \multicolumn{3}{c|}{\textbf{Preprocess Time (s)}} & \multicolumn{3}{c}{\textbf{Solve Time (s)}} \\
& cache & no cache & no inc & cache & no cache & no inc \\ \hline
multidim\_32b & 0.153 & 2.96 & 99.17 & 3.54 & 4.61 & 31.84 \\
multidim\_nested & 0.154 & 2.94 & 103.33 & 3.67 & 4.47 & 32.91 \\
multidim\_47b & 0.178 & 2.90 & 134.51 & 4.64 & 5.53 & 48.10 \\
unique\_simple & 0.137 & 5.81 & 98.13 & 5.63 & 7.68 & 57.76 \\
unique\_nested & 0.137 & 5.69 & 94.79 & 5.16 & 7.23 & 55.45 \\
bit\_count & 0.017 & 0.120 & 8.22 & 0.319 & 0.339 & 15.65 \\
mem\_alloc & 0.004 & 0.132 & 2.61 & 0.187 & 0.207 & 1.02 \\
clog & 0.002 & 0.117 & 854.3 & 0.130 & 0.140 & 1.15 \\
hash\_inverse & 0.001 & 0.094 & 0.006 & 0.070 & 0.071 & 2.70 \\

float\_compare & 0.005 & 0.126 & 2.84 & 0.118 & 0.125 & 22.70 \\
conditional\_a & 0.002 & 0.142 & 98.7 & 0.073 & 0.082 & 1.00 \\
conditional\_b & 0.136 & 0.178 & 564.0 & 0.110 & 0.112 & 0.30 \\
ite & 2.18 & 2.31 & 50.82 & 1.42 & 1.53 & 3.56 \\
subset\_sum & 0.004 & 0.860 & 4.39 & 0.552 & 0.589 & 89.70 \\
solve\_priority & 2.80 & 4.16 & 1984.15 & 18.70 & 19.14 & 512.17 \\
\hline
\end{tabular}
\end{table}

Table~\ref{tab:detailed_performance_reordered} profiles preprocessing and solving time (including bit-blasting, CNF encoding, and SAT solving) under three configurations: no incrementality (\emph{no inc}), incremental solving without assumption caching (\emph{no cache}), and incremental solving with caching (\emph{cache}). Without incrementality, \texttt{randomize()} call repeats full preprocessing and solving, leading to a dominant cost. Enabling incrementality reduces preprocessing to assumption handling after the initial solve, while caching further eliminates this overhead when assumptions remain unchanged. In the solve phase, incremental solving prevents redoing bit-blasting and  encoding, allows reuse of learned clauses, and caching avoids repeated term registration for stable assumptions. As a result, the configuration with both incrementality and caching consistently achieves the lowest preprocessing and solve times, whereas disabling incrementality incurs higher overhead in both phases.

\section{Conclusion}
In this tool paper, we formalized the Dynamic Data Structure Constraint Satisfaction Problem and introduced a dependency-aware, incremental solving framework tailored to constrained-random verification. The framework reduces redundant work while maintaining scalability, enabling dynamic data structures to be handled efficiently across solver invocations. As for future work, we plan to explore tool support for more complex mechanisms such as soft constraints, user-defined distributions, and conditionally activated constraint branches.



\begin{thebibliography}{8}
\bibitem{bitwuzla2023}
Niemetz, A., Preiner, M., Biere, A.:
Bitwuzla: A modern satisfiability modulo theories solver.
In: Gurfinkel, A., Heule, M. (eds.)
Computer Aided Verification -- CAV 2023.
LNCS, vol. 13965, pp. 3--17.
Springer, Cham (2023).
\href{https://doi.org/10.1007/978-3-031-37703-7_1}{https://doi.org/10.1007/978-3-031-37703-7\_1}

\bibitem{mehta2018}
Mehta, A.B.:
ASIC/SoC Functional Design Verification:
A Comprehensive Guide to Technologies and Methodologies.
Springer, Cham (2018).
\bibitem{cohen2023}
Cohen, B.:
Dynamic Data Structures in Assertions.
Technical report, SystemVerilog.us (2023).
\href{https://systemverilog.us/vf/SVA_aa_q_V1030b.pdf}{https://systemverilog.us/vf/SVA\_aa\_q\_V1030b.pdf}

\bibitem{verfaillie2005}
Verfaillie, G., Jussien, N.:
Constraint solving in uncertain and dynamic environments: A survey.
Constraints 10, 253--281 (2005).
\href{https://doi.org/10.1007/s10601-005-2239-9}{https://doi.org/10.1007/s10601-005-2239-9}

\bibitem{schiex1995}
Schiex, T., Verfaillie, G.:
Maintien de solution dans les problemes dynamiques de satisfaction de contraintes: bilan de quelques approches.
Revue d'intelligence artificielle 9, 269--309 (1995).

\bibitem{verfaillieschiex1994b}
Verfaillie, G., Schiex, T.:
Dynamic backtracking for dynamic constraint satisfaction problems.
In: ECAI-94 Workshop on Constraint Satisfaction Issues Raised by Practical Applications, pp. 1--8 (1994).

\bibitem{schiexverfaillie1994a}
Schiex, T., Verfaillie, G.:
Nogood recording for static and dynamic constraint satisfaction problems.
International Journal of Artificial Intelligence Tools 3(2), 187--207 (1994).
\href{https://doi.org/10.1142/S0218213094000108}{https://doi.org/10.1142/S0218213094000108}

\bibitem{schiexverfaillie1994b}
Schiex, T., Verfaillie, G.:
Stubbornness: A possible enhancement for backjumping and nogood recording.
In: Proceedings of the 11th European Conference on Artificial Intelligence (ECAI 1994),
pp. 165--172 (1994).

\bibitem{ridgeway2012svcr}
Ridgeway, J.:
The Top Most Common SystemVerilog Constrained Random Gotchas.
In: DVCon (Design \& Verification Conference) (2012).




\bibitem{bessiere1991arc}
Bessière, C.:
Arc-consistency in dynamic constraint satisfaction problems.
In: Proceedings of the 9th National Conference on Artificial Intelligence (AAAI-91),
pp. 221--226 (1991).

\bibitem{amilhastre2002}
Amilhastre, J., Fargier, H., Marquis, P.:
Consistency restoration and explanations in dynamic constraint satisfaction problems.
Artificial Intelligence 135(1--2), 199--234 (2002).

\bibitem{improvingglucose2016}
Audemard, G., Lagniez, J.-M., Simon, L.:
Improving Glucose for incremental SAT solving with assumptions: Application to MUS extraction.
In: Järvisalo, M., Van Gelder, A. (eds.)
Theory and Applications of Satisfiability Testing -- SAT 2013.
LNCS, vol. 7962, pp. 309--317.
Springer, Berlin, Heidelberg (2013).
\href{https://doi.org/10.1007/978-3-642-39071-5_23}{https://doi.org/10.1007/978-3-642-39071-5\_23}

\bibitem{cimatti2012}
Nadel, A., Ryvchin, V.:
Efficient SAT solving under assumptions.
In: Cimatti, A., Sebastiani, R. (eds.)
Theory and Applications of Satisfiability Testing -- SAT 2012.
LNCS, vol. 7317, pp. 242--255.
Springer, Berlin, Heidelberg (2012).
\href{https://doi.org/10.1007/978-3-642-31612-8_19}{https://doi.org/10.1007/978-3-642-31612-8\_19}

\bibitem{bofill2009}
Bofill, M., Suy, J., Villaret, M.:
A system for solving constraint satisfaction problems with SMT.
In: Strichman, O., Szeider, S. (eds.)
Theory and Applications of Satisfiability Testing -- SAT 2010.
LNCS, vol. 6175, pp. 300--305.
Springer, Berlin, Heidelberg (2010).
\href{https://doi.org/10.1007/978-3-642-14186-7_25}{https://doi.org/10.1007/978-3-642-14186-7\_25}

\bibitem{petke2015}
Petke, J.:
Bridging Constraint Satisfaction and Boolean Satisfiability.
Springer, Cham (2015).
\href{https://doi.org/10.1007/978-3-319-21810-6}{https://doi.org/10.1007/978-3-319-21810-6}

\bibitem{molina2007fv}
Molina, A., Cadenas, O.:
Functional verification: Approaches and challenges.
Latin American Applied Research 37(1), 65--69 (2007).
\bibitem{ludden2002power4}
Ludden, J.M., Roesner, W., Heiling, G.M., Reysa, J.R., et al.:
Functional verification of the POWER4 microprocessor and POWER4 multiprocessor systems.
IBM Journal of Research and Development 46(1), 53--76 (2002).
\href{https://doi.org/10.1147/rd.461.0053}{https://doi.org/10.1147/rd.461.0053}

\bibitem{foster2016wrg}
Foster, H.D.:
Trends in Functional Verification: A 2016 Industry Study.
In: DVCon (Design \& Verification Conference) (2016).

\bibitem{tasiran2001coverage}
Tasiran, S., Keutzer, K.:
Coverage metrics for functional validation of hardware designs.
IEEE Design \& Test of Computers 18(4), 36--45 (2001).
\href{https://doi.org/10.1109/54.936247}{https://doi.org/10.1109/54.936247}

\bibitem{jayasena2023directed}
Jayasena, A., Mishra, P.:
Directed test generation for hardware validation: A survey.
ACM Computing Surveys 56(5), Article 132, 1--36 (2024).
\href{https://doi.org/10.1145/3637868}{https://doi.org/10.1145/3637868}

\bibitem{teplitsky2013coverage}
Teplitsky, M.:
Coverage-driven distribution of constrained-random stimuli.
In: DVCon (Design \& Verification Conference) (2013).

\bibitem{brailsford1999csp}
Brailsford, S.C., Potts, C.N., Smith, B.M.:
Constraint satisfaction problems: Algorithms and applications.
European Journal of Operational Research 119(3), 557--581 (1999).
\href{https://doi.org/10.1016/S0377-2217(98)00364-6}{https://doi.org/10.1016/S0377-2217(98)00364-6}

\bibitem{wu2013framework}
Wu, B.-H., Huang, C.-Y. (Ric):
A robust constraint solving framework for multiple constraint sets in constrained-random verification.
In: DAC, pp. 119:1--119:7 (2013).
\href{https://doi.org/10.1145/2463209.2488880}{https://doi.org/10.1145/2463209.2488880}
\bibitem{chen2013reuse}
Chen, W., Wang, L.-C., Bhadra, J., Abadir, M.S.:
Simulation knowledge extraction and reuse in constrained random processor verification.
In: DAC, pp. 120:1--120:6 (2013).
\href{https://doi.org/10.1145/2463209.2488881}{https://doi.org/10.1145/2463209.2488881}

\bibitem{shacham2008relaxedscoreboard}
Shacham, O., Wachs, M., Solomatnikov, A., Firoozshahian, A., Richardson, S., Horowitz, M.:
Verification of Chip Multiprocessor Memory Systems Using a Relaxed Scoreboard.
In: MICRO 2008: Proceedings of the 41st Annual IEEE/ACM International Symposium on Microarchitecture,
pp. 294--305 (2008).
\href{https://doi.org/10.1109/MICRO.2008.4771799}{https://doi.org/10.1109/MICRO.2008.4771799}

\bibitem{freitas2013concurrent}
Freitas, L.S., Rambo, E.A., dos Santos, L.C.V.:
On-the-fly Verification of Memory Consistency with Concurrent Relaxed Scoreboards.
In: DATE 2013, pp. 631--636 (2013).
\href{https://doi.org/10.7873/DATE.2013.138}{https://doi.org/10.7873/DATE.2013.138}

\bibitem{mittal1990dcsp}
Mittal, S., Falkenhainer, B.:
Dynamic constraint satisfaction problems.
In: AAAI-90 Proceedings, pp. 25--32 (1990).
\href{https://cdn.aaai.org/AAAI/1990/AAAI90-004.pdf}{https://cdn.aaai.org/AAAI/1990/AAAI90-004.pdf}

\bibitem{wallace2009dcsp}
Wallace, R.J., Grimes, D., Freuder, E.C.:
Solving dynamic constraint satisfaction problems by identifying stable features.
In: IJCAI, pp. 621--627 (2009).
\bibitem{davidson2020cp2smt}
Davidson, E., Akg\"un, \"O., Espasa, J., Nightingale, P.:
Effective Encodings of Constraint Programming Models to SMT.
In: Simonis, H. (ed.) Principles and Practice of Constraint Programming (CP 2020),
LNCS, vol. 12333, pp. 143--159.
Springer, Cham (2020).
\href{https://doi.org/10.1007/978-3-030-58475-7_9}{https://doi.org/10.1007/978-3-030-58475-7\_9}

\bibitem{bofill2020rcpsp}
Bofill, M., Coll, J., Suy, J., Villaret, M.:
SMT encodings for Resource-Constrained Project Scheduling Problems.
Computers \& Industrial Engineering 149, 106777 (2020).
\href{https://doi.org/10.1016/j.cie.2020.106777}{https://doi.org/10.1016/j.cie.2020.106777}

\bibitem{synopsysvcsT2022ucli}
Synopsys, Inc.:
VCS Unified Command Line Interface User Guide.
Version T-2022.06 (2022).
\href{https://spdocs.synopsys.com/dow_retrieve/qsc-t/vg/VCS/T-2022.06/PDFs/vcs.pdf}{https://spdocs.synopsys.com/dow\_retrieve/qsc-t/vg/VCS/T-2022.06/PDFs/vcs.pdf}

\bibitem{primariusverisim_web}
Primarius Technologies Co., Ltd.:
VeriSim -- Digital design EDA.
Available at:
\href{https://www.primarius-tech.com/en/products/digital_design_eda/VeriSim-EN.html}{https://www.primarius-tech.com/en/products/digital\_design\_eda/VeriSim-EN.html}

\bibitem{song2021learning}
Song, W., Cao, Z., Zhang, J., Lim, A.:
Learning variable ordering heuristics for solving constraint satisfaction problems.
Engineering Applications of Artificial Intelligence 109, 104603 (2022).
\href{https://doi.org/10.1016/j.engappai.2021.104603}{https://doi.org/10.1016/j.engappai.2021.104603}

\bibitem{marx2009hypergraph}
Marx, D.:
Tractable hypergraph properties for constraint satisfaction and conjunctive queries.
Journal of the ACM 60(6), Article 42, 42:1--42:51 (2013).
\href{https://doi.org/10.1145/2535926}{https://doi.org/10.1145/2535926}

\bibitem{bembenek2020datalog}
Bembenek, A., Ballantyne, M., Greenberg, M., Amin, N.:
Datalog-Based Systems Can Use Incremental SMT Solving (Extended Abstract).
In: International Conference on Logic Programming (ICLP) Technical Communications.
EPTCS 325, 18--20 (2020).
\href{https://doi.org/10.4204/EPTCS.325.7}{https://doi.org/10.4204/EPTCS.325.7}



\end{thebibliography}
\end{document}